\documentclass[aps,twocolumn,floats,groupedaddress,bibnotes]{revtex4b1}
\usepackage{graphicx,graphics}

\begin{document}
\title{Universal Selftrapping in Nonlinear Tight-binding Lattices}
\author{C.A. Bustamante}
\email[]{carlosb@macul.ciencias.uchile.cl}
\author{M.I. Molina}
\affiliation{Facultad de Ciencias, Departamento de F\'{\i}sica, Universidad de Chile \\ Casilla 653, Las Palmeras 3425, Santiago, Chile.}

\begin{abstract}
  We show that nonlinear tight-binding lattices of different
  geometries and dimensionalities, display an universal selftrapping
  behavior.  First, we consider the single nonlinear impurity problem
  in various tight-binding lattices, and use the Green's function
  formalism for an exact calculation of the minimum nonlinearity
  strength to form a stationary bound state. For all lattices, we find
  that this critical nonlinearity parameter (scaled by the energy of
  the bound state), in terms of the nonlinearity exponent, falls
  inside a narrow band, which converges to $e^{1/2}$ at large exponent
  values.  Then, we use the Discrete Nonlinear Schr\"{o}dinger (DNLS)
  equation to examine the selftrapping dynamics of a single
  excitation, initially localized on the single nonlinear site, and
  compute the critical nonlinearity parameter for abrupt dynamical
  selftrapping. For a given nonlinearity exponent, this critical
  nonlinearity, properly scaled, is found to be nearly the same for
  all lattices.  Same results are obtained when generalizing to
  completely nonlinear lattices, suggesting an underlying selftrapping
  universality behavior for all nonlinear (even disordered)
  tight-binding lattices described by DNLS.
\end{abstract}
\pacs{71.23.An,\ 71.55.-i}
\date{\today}
\maketitle

The Discrete Nonlinear Schr\"{o}dinger (DNLS) equation is a
paradigmatic equation describing among others, dynamics of polarons in
deformable media\cite{davidov}, local modes in molecular
systems\cite{bishop} and power exchange among nonlinear coherent
couplers in nonlinear optics\cite{optics}. Its most striking feature
is the possibility of ``selftrapping'', that is, the clustering of
vibrational energy or electronic probability or electromagnetic energy
in a small region of space. In a condensed matter context, the DNLS
equation has the form
\begin{equation}
i {d\; C_{\bf n}\over{d\; t}} = \epsilon_{\bf n}\ C_{\bf n} + V\; {\sum_{{\bf m}}}^{'}\; C_{\bf m} -
\chi_{{\bf n}}\; | C_{\bf n} |^{\alpha}\; C_{\bf n} 
\label{eq:dnls}
\end{equation}
where $C_{\bf n}$ is the probability amplitude of finding the electron
(or excitation) on site {\bf n} of a $d$-dimensional lattice,
$\epsilon_{\bf n}$ is the on--site energy, $V$ is the transfer matrix
element, $\chi_{\bf n}$ is the nonlinearity parameter at site ${\bf
  n}$ and $\alpha$ is the nonlinearity exponent.  The prime in the sum
in (\ref{eq:dnls}) restricts the summation to nearest--neighbors only.
In the {\em conventional} DNLS case, $\alpha = 2$ and $\chi_{\bf n}$
is proportional to the square of the electron-phonon coupling at site
${\bf n}$.\cite{chi}

Considerable work has been carried out in recent years to understand
the stationary and dynamical properties of Eq. (\ref{eq:dnls}) in
various cases. In particular, we point out the studies on the
stability of the stationary solutions in one and two dimensions for
the homogeneous case ($\epsilon_{\bf n} =0, \chi_{\bf n} =
\chi$)\cite{laedke1,laedke2}, the effect of point linear impurities on
the stability of the 2-D DNLS solitons\cite{christiansen}, the effects
of nonlinear disorder ($\epsilon_{\bf n} = 0, \chi_{\bf n}$
random)\cite{mt_prl} and of linear disorder ($\chi_{\bf n} = \chi,
\epsilon_{\bf n}$ random)\cite{mm_prb98} on the selftrapping dynamics
of initially localized and extended excitations in a chain. The
results obtained in these studies suggest that, in general, the effect
of nonlinearity is quite {\em local} for initially localized
excitations, and that disorder leaves the narrow selftrapped
excitations unaffected, although it does affect the propagation of the
untrapped portion (``radiation''). In this Letter we show that, for an
initially localized excitation, the dynamics of selftrapping in
various different lattices of different dimensionalities, is universal
and depends mainly on the nonlinearity strength at the initial site,
the nonlinearity exponent and the coordination number, and much more
weakly on other topological features of the lattice.

\bigskip
\noindent
{\em Bound states}.\ \ A tight correlation has been observed between
the existence of bound states for a given nonlinear lattice and the
ability of the lattice to selftrap an initially completely--localized
excitation:\ the critical nonlinearity strength for dynamical
selftrapping is always greater than the one needed to produce bound
state(s). We begin by showing that the minimum nonlinearity needed to
produce a bound state in different lattices, shows universal features.

We consider the problem of determining the bound state for an electron
in a $d$--dimensional homogeneous lattice that contains a single
generalized nonlinear impurity at the origin ${\bf n = 0}$.  The
Hamiltonian is ${\tilde{H}}={\tilde{H}}_0+{\tilde{H}}_1$, where
${\tilde{H}}_0=V\; \sum_{n.n}(\; |{\bf{n}}\rangle\langle{\bf{m}}| +
\mbox{h.c.}\; )$ is the unperturbed tight--binding Hamiltonian with
hopping constant $V$ and ${\tilde{H}}_1 = \chi |C_{\bf 0}|^\alpha\;
|\bf{0}\rangle\langle\bf{0}|$ corresponds to the nonlinear impurity
perturbation.  The $\{|\bf{n}\rangle\}$ represent Wannier electronic
states, and we have set $\epsilon_{\bf n} = 0$. For convenience we
normalize all energies to a half bandwidth, $B$ and define: $z\equiv
E/B$, $H\equiv {\tilde{H}}/B$ and $\gamma \equiv \chi /B$. The
dimensionless lattice Green function $G=1/(z-H)$ can be formally
expanded as\cite{economou}
$G=G^{(0)}+G^{(0)}H_1G^{(0)}+G^{(0)}H_1G^{(0)}H_1G^{(0)}+...$, where
$G^{(0)}$ is the unperturbed ($\gamma =0$) Green function and
$H_1=\gamma |C_{\bf{0}}|^\alpha \; |{\bf{0}}\!><~\!\!{\bf{0}}|$. The
sum can be carried out exactly to yield
\begin{equation}
G_{{\bf m n}}=G_{{\bf m n}}^{(0)}+{\frac{\gamma |C_{\bf 0}|^\alpha \;
G_{{\bf m 0}}^{(0)}\; G_{{\bf 0 n}}^{(0)}}{{%
1-\gamma |C_{\bf 0}|^\alpha \; G_{{\bf 0 0}}^{(0)}}}}.  \label{eq:gtotal}
\end{equation}
where $G_{{\bf m n}} = \langle {\bf m}| G |{\bf n} \rangle$.
The energy of the bound state(s), $z_b$ is obtained
from the poles of $G_{{\bf m n}} $, {\em i.e.}, by solving $1 =
\gamma |C_{{\bf 0}}^{(b)}|^\alpha \; G_{{\bf{0 0}}}^{(0)}$.
The bound state amplitudes $C_{\bf n}^{(b)}$ are obtained from the
residues of $G_{\bf m n}(z)$ at $z = z_b$. In particular,  
$|C_{\bf 0}^{(b)}|^2 = Res\{G_{0 0}(z)\}_{z=z_b} =
- {G_{\bf 0 0}^{(0)}}^{2} (z_{b}) / G_{\bf 0 0}^{'(0) } (z_{b})$.
Inserting this into the bound state energy equation leads to
\begin{equation}
1 = {\gamma {G_{\bf 0 0}^{(0)}}^{\;\alpha + 1} (z_{b})\over{[-
G_{\bf 0 0 }^{'(0)} (z_{b})\; ]^{\;\alpha/2}}}.\label{eq:zb}
\end{equation}

We proceed to solve (\ref{eq:zb}) numerically, using the exact, known
expressions for $G_{\bf 0 0}^{(0)}$ for several
lattices\cite{economou, green}:\ one-dimensional (1-D), square,
triangular, simple cubic and Bethe lattices with connectivities $3$,
$5$ and $100$. This allows us to compare lattices with different
dimensionality, coordination number $Z$, length of shortest loops,
etc. In general, for a given $\alpha$ value there will be a minimum
value of $\chi$ below (above) which, there is (are) no (two) bound
state(s). Just at the critical nonlinearity value, we obtain exactly
one bound state. The exception is the 1-D lattice where one needs in
addition, $\alpha \geq 2$\cite{mth_pre}.

 \begin{figure}[tb]
   \begin{center}
     \includegraphics[width=.5 \textwidth]{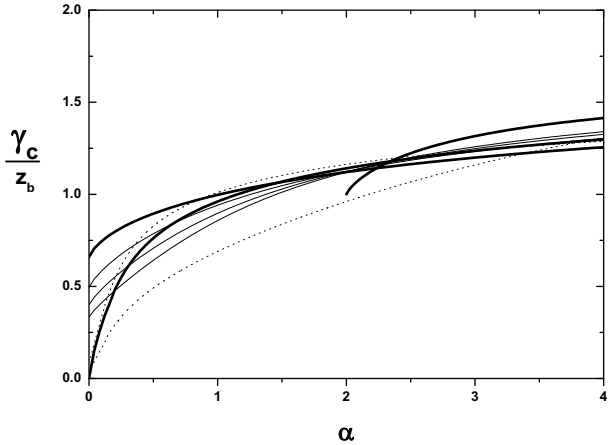}
     \caption{The critical nonlinear paremeter, $\gamma_c/z_b$, for Bound States. Thick lines correspond to Cubic, Square and 1D cases. Thin lines correspond to Bethe lattices with $K= 3, 5$ and 100 in ascending order near $\alpha =0$. Dotted lines represent the Triangular cases: $sgn(\chi/V)>0$ (upper line) and $sgn(\chi/V)<0$ }
     \label{fig:bs}
   \end{center}
 \end{figure}

Figure 1 shows the critical nonlinearity parameter $\gamma_{c}$,
scaled by the energy of the bound state, in terms of $\alpha$, the
nonlinearity exponent, for all the lattices examined. These curves are
independent of $sgn(\chi/V)$, except for the triangular lattice, due
to the asymmetry of its Green function with respect to the energy
variable. In this case there are two curves depending on
$sgn(\chi/V)$.  All curves in Fig.1 fall inside a ``band'' which
narrows as $\alpha$ increases, converging towards a constant value. To
calculate it, we solve (\ref{eq:zb}) exactly in two cases: the
one-dimensional lattice\cite{mth_pre} and the Bethe lattice in the
limit of infinite connectivity (numerically indistinguishable from $K
= 100$). In both cases we obtain:
\begin{equation}
\lim_{\alpha\rightarrow \infty} \left({\gamma_{c}\over{z_{b}}}\right)  = e^{1/2} \sim 1.65.\label{eq:gc}
\end{equation}
We have traced the validity of (\ref{eq:gc}) for the other lattices up
to high $\alpha$ values ($10^3$ for the square and cubic lattices;
$10^{5}$ for the rest) with no discernible deviation.

\bigskip
\noindent
{\em Selftrapping Dynamics}.\ \ We now examine the ability of a given
lattice to dynamically selftrap an excitation, originally placed
completely on the impurity site, by computing the minimum nonlinearity
needed to give rise to abrupt selfptrapping. The time evolution is
given by Eq.(\ref{eq:dnls}) with $\epsilon_{\bf n} = 0$ and $\chi_{\bf
  n} = \chi\ \delta_{{\bf n},{\bf 0}}$. The numerical scheme is that
of a fourth-order Runge-Kutta, where the accuracy is monitored through
total probability conservation. To avoid undesired boundary effects, a
self-expanding lattice is used\cite{mt_prl}. To ascertain the presence
or absence of a sharp selftrapping transition, we compute the
long-time average probability at the impurity site, defined by
\begin{equation}
P_{\bf 0} =
\lim_{T\rightarrow \infty} (1/T)\ \int_{0}^{T} | C_{\bf 0}(t) |^{2} dt, \hspace{1cm}
| C_{\bf 0}(0) | = 1.
\end{equation}
Typically, $P_{\bf 0}$ vanishes for nonlinearity parameters below a
critical value $\chi_{c}$ and the particle escapes from the impurity
site in a ballistic manner. This is determined from an examination of
the excitation's mean square displacement $\langle u(t) \rangle =
\sum_{\bf n} {\bf n}^{2}\ | C_{\bf n} |^{2}$. For nonlinearity values
greater than $\chi_{c}$, $P_{\bf 0}$ remains finite and increases with
$\chi$, converging towards unity at large $\chi$. The untrapped
portion escapes to infinity, also in a ballistic manner, but with a
much lower ``speed'' $\sqrt{\langle u(t) \rangle} / Vt$. Thus, from
the examination of $P_{\bf 0}$ we determine the critical nonlinearity
parameter $\chi_{c}$ for dynamical selftrapping (usually for $P_{\bf
  0}\approx 1/2 $).

For a particular lattice and a given exponent $\alpha$, we numerically
determine the critical nonlinear parameter $\chi_{c}$, scaled by
$E_{b}$ (where $E_{b}$ is the unnormalized bound state energy
correspondig to this $\chi_c$) for abrupt selftrapping.  Figure 2
shows $\chi_{c}/E_{b}$ for all the lattices examined, and for several
$\alpha$ values that give rise to sharp selftrapping (for $\alpha <
1$, the selftrapping is not sharp). We see that, for the wide range of
geometries and dimensionalities involved, this critical (dynamical)
nonlinearity is nearly independent of the lattice and increases
monotonically with the nonlinearity exponent. This is specially true
in the all--important {\em conventional} DNLS case ($\alpha = 2$).  It
would seem that, in the $\alpha$ regime where abrupt selftrapping
takes place ($\alpha \geq 1$), the only relevant parameters are the
nonlinearity at the impurity site and the coordination number of the
lattice. The rest of the topological features is of secondary
importance.  In all cases, with the exception of the triangular
lattice, the critical nonlinearity is independent of the sign of $\chi
/V$. For the triangular lattice we note that $\chi_{c}/E_b$ gets
shifted a bit upon changing the sign of $\chi /V$. This probably
trails back to the asymmetry of the unperturbed triangular lattice's
Green function $G_{{\bf{0 0}}}^{(0)}$ under a sign change of its
argument\cite{green}. All the rest of the lattices are symmetric in
that respect. The increase of $\chi_{c}$ with $\alpha$ is to be
expected since, in the continuum limit , increasing $\alpha$ is
equivalent to increasing the dimensionality of the
system\cite{laedke1,juul}; this in turns increases the effective
coordination number making it harder to selftrap the excitation;
hence, the need for larger nonlinearities. Also, the obtained values
of $\chi_{c}$ in the dynamical case are all higher than for the bound
state case, confirming the conjecture that the onset of the stationary
bound state is a precursor for dynamical selftrapping.  However, the
lack of a superposition principle, makes it hard to establish formally
the (observed) connection between the dynamical and the stationary
DNLS problem. An alternative normalization for $\chi_{c}$ is to use
the half bandwidth $B$ instead of $E_{b}$. In that case, all the
curves in figure 2 lose a bit of flatness, but the tendency is
otherwise unaltered.

 \begin{figure}[htb]
   \begin{center}
     \includegraphics[width=.5 \textwidth]{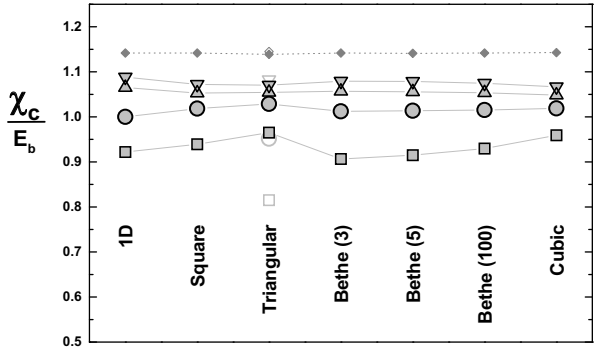}
     \caption{The dynamic critical nonlinearity parameter $\chi_c$ scaled by the bound state energy for one nonlinear impurity in  various lattices. The values for the nonlinearity exponent are $\alpha=1,2,3,4$ from bottom to top (hollow symbols represent the case $sgn(\chi/V)<0$). The limiting (upper) curve correspond to $\alpha=1000$.}
     \label{fig:impurity}
   \end{center}
 \end{figure}

\noindent
We now recompute all of the above selftrapping dynamics calculations,
this time using completely nonlinear lattices ($ \epsilon_{\bf n} = 0,
\chi_{\bf n} = \chi$) and same initial conditions ($C_{\bf n} =
\delta_{\bf n\ 0}$). Figure 3 shows the results obtained for the
critical nonlinearities. The curves are virtually the same as the ones
in Fig.2. (The case $\alpha = 1$ does not display abrupt selftrapping
like the rest, thus $\chi_c$ is not precisely defined here). This is
due to the fact that, once the abrupt selftrapping is set, most of the
probability is on the initial site, which gives, by conservation of
probability, very small probability amplitudes for the rest of the
lattice sites, making their nonlinear contribution negligible: they
have become, in fact, linear for all selftrapping purposes and, in
this way we are back to the single nonlinear impurity results. The
greater the $\alpha$ value, the closer the system to the nonlinear
impurity case. This is vividly illustrated by the limiting curves for
large $\alpha$ in Figs. 2 and 3., which coincide.

 \begin{figure}[htb]
   \begin{center}
     \includegraphics[width=.5 \textwidth]{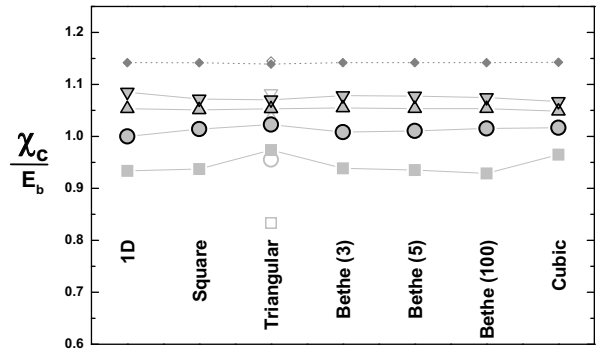}
     \caption{The dynamic critical nonlinearity parameter scaled by the bound state energy (from the stationary impurity problem) for completely nonlinear lattices. The values for the nonlinearity exponent are $\alpha=1,2,3,4$ from bottom to top (the case $\alpha = 1$ does not show abrupt selftrapping; hollow symbols represent $sgn(\chi /V)<0$). The limiting (upper) curve correspond to $\alpha=1000$.  }
     \label{fig:NL}
   \end{center}
 \end{figure}

\noindent
This characterization of the selftrapping properties of nonlinear tight--binding
lattices of different geometries and dimensionalities, in terms of a single
parameter, namely the bound state energy for the one--impurity problem
(or the half bandwidth B for quick estimations), could be useful in several
areas, given the paradigmatic character of DNLS.

\bigskip
\begin{center}
  ACKNOWLEDGMENTS
\end{center}

 This work was partially supported by FONDECYT, proyects 1990960,
 2980033 and 4990004.

\printfigures

\end{document}